\documentclass[12pt]{article}
\usepackage{latexsym,amsmath,amssymb, amsfonts,mathrsfs}

\newcommand{\sect}[1]{\setcounter{equation}{0}\section{#1}}

\newcommand{\eq}{\begin{equation}}
\newcommand{\eqa}{\begin{eqnarray}}
\newcommand{\en}{\end{equation}}
\newcommand{\ena}{\end{eqnarray}}
\newcommand{\enn}{\nonumber \end{equation}}

\def\sk{\vskip .4cm}
\def\noi{\noindent}
\def\om{\omega}
\def\al{\alpha}

\def\ga{\gamma}

\def\la{\lambda}

\def\epsi{\varepsilon}
\def\we{\wedge}

\def\de{\delta}

\def\noi{\noindent}

\def\epsi{\varepsilon}

\def\we{\wedge}

\def\de{\delta}

\def\st{\star}

\def\Phi{\phi}

\def\westar{\we_\star}

\def\omtilde{\tilde \om}
\def\Vtilde{\tilde V}

\def\rtilde{\tilde r}
\def\epsitilde{\tilde \epsi}

\def\psibar{\bar \psi}

\def\Om{\Omega}

\def\A{\cal A}
\def\of{\bar{\rm f}}

\def\dd{{\rm d}}

\def\FF{\mathcal F}

\begin{document}

\begin{titlepage}
\begin{center}{\Large \bf Extended gravity from
    noncommutativity}
\\[3em]
{\large {\bf Paolo Aschieri\footnote{\normalsize{The results presented in this
      Proceedings are based on joint work with Leonardo Castellani.}}}}
\\ [1em] {\sl Dipartimento di Scienze e Innovazione Tecnologica
\\ INFN Gruppo collegato di Alessandria,\\Universit\`a del Piemonte
Orientale,\\ Viale T. Michel 11,  15121 Alessandria, Italy}\\
 {\small aschieri@to.infn.it}
\end{center}

\begin{abstract}

\vskip 0.2cm

We review the first order theory of gravity (vierbein formulation)  
on noncommutative spacetime studied in \cite{AC1, AC2}. The first order formalism allows 
to couple the theory to fermions. This NC action is then
reinterpreted (using the Seiberg-Witten map) 
as a gravity theory on commutative spacetime that contains terms with higher
derivatives and higher powers of the curvature and depend on the
 noncommutativity parameter $\theta$.
  When the
noncommutativity is switched off we recover the 
 usual gravity action coupled to fermions.  

The first nontrival
corrections to the usual gravity action coupled to fermions  
are presented in a manifest Lorentz invariant form.

 \end{abstract}

\end{titlepage}

\newpage
\section{Introduction}
In the passage from classical 
mechanics to quantum mechanics classical observables become noncommutative. 
Similarly we expect that in the passage from classical gravity to quantum 
gravity, gravity observables, i.e. spacetime itself, with its coordinates 
and metric structure, will become noncommutative. Thus by formulating Einstein 
gravity on noncommutative spacetime we may learn some aspects of  quantum gravity.

Planck scale noncommutativity is further supported by Gedanken experiments 
that aim at probing spacetime structure at very small distances. 
They show that due to gravitational backreaction one cannot test 
spacetime at those distances. For example, 
in relativistic quantum mechanics the position of a particle can
be detected with a precision at most of the order of its 
Compton wave length $\lambda_C=\hbar/mc$.
Probing spacetime at infinitesimal distances implies an extremely 
heavy particle that in turn curves spacetime itself. 
When $\lambda_C$ is of the order of the Planck length, the spacetime
curvature radius due to the particle has the same order of magnitude and the 
attempt to measure spacetime structure beyond Planck scale fails.

This Gedanken experiment supports finite 
reductionism. 
It  shows that the description of spacetime
as a continuum of points (a smooth manifold) is an 
assumption no more justified at Planck scale. 
It is then natural to 
relax this assumption and conceive a noncommutative spacetime, 
where uncertainty relations and discretization naturally arise. 
In this way the dynamical feature of spacetime that prevents from testing 
sub-Plankian scales is explained by incorporating it at a deeper kinematic level.
A similar mechanism happens for example in the passage from 
Galilean to special relativity. Contraction of distances and 
time dilatation can be explained in Galilean relativity: 
they are a consequence of the interaction between ether and the 
body in motion. In special relativity they become a kinematic feature.
\sk
The noncommutative gravity theory we present following \cite{AC1,AC2} is an effective theory
that may capture some aspects of a quantum gravity theory. Furthermore
we reinterpret space-time noncommutativity  as extra
interaction terms on commutative spacetime, in this way the theory 
 is equivalent to a higher derivative and curvature  extension
of Einstein general relativity. 
We have argued that spacetime noncommutativity
should be relevant at Planck scale, however the physical phenomena it
induces can also appear at larger scales. For example, due to
inflation, noncommutativity of spacetime at inflation scale (that may
be as low as Planck scale) can affect cosmological perturbations 
and possibly the cosmic microwave
background spectrum; see
for example \cite{Lizzi:2002ib}. 
We cannot exclude that this noncommutative extension 
of gravity can be relevant for advancing in our understanding of 
nowadays open questions in cosmology.

\sk
In this contribution, after a short overview of possible noncommutative
approaches, we outline the Drinfeld twist approach and review the geometric
formulation of theories on noncommutative spacetime \cite{book}. This allows to
construct actions invariant under diffeomorphisms. 
In section 4 we first present usual gravity coupled to fermions in an index free
formalism suited for its generalization to the noncommutative case.
Then we discuss gauge theories on noncommutative space and in
particular local Lorentz symmetry
($SO(3,1)$-gauge symmetry), indeed we need a vierbein formulation of
noncommutative gravity in order to couple gravity to spinor
fields. The noncommutative Lagrangian coupled to spinor fields is 
then presented.
In section 5 we reinterpret this NC gravity as an extended
gravity theory on commutative spacetime. This is done via the Seiberg-Witten map from noncommutative to commutative gauge fields. 
The resulting gravity theory then depends on the usual gravitational degrees of freedom
plus the noncommutative degrees of freedom, these latter are encoded in a set of
mutually commuting vector fields $\{X_I\}$. The leading correction terms
to the usual action are explicitly
calculated in section 6. They couple  spinor fields and their covariant derivatives
to derivatives of the curvature tensor and of the vierbein.
It is interesting to consider a kinetic term for these vector
fields, so that the noncommutative structure  of spacetime, as well as
its metric structure depend on the matter content of spacetime itself.
A model of dynamical noncommutativity is presented in \cite{ACpreparation}. 
Noncommutative vierbein gravity can also be coupled to scalar
fields \cite{ACpreparation} and to gauge fields \cite{AC4} .

\section{NC geometry approaches}
Before entering the details of the theory, we briefly  frame it in
the context of noncommutative geometry approaches.

The easiest way to describe a noncommutative spacetime is via the
noncommutative algebra of its coordinates, i.e., we give a set of
generators and relations.
  For example
\begin{eqnarray}
\label{uuno}&[ x^\mu,  x^\nu]=i\theta^{\mu\nu}   ~~~~~~~~&{\mbox{\sl canonical}}\\[1em]
\label{ddue}&~~~~~[ x^\mu, x^\nu]=if^{\mu\nu}_{~~\sigma} x^\sigma   ~~~~~~~~&{\mbox{\sl Lie algebra}}
\\[1em]
\label{ttre}&~~~~~ x^\mu  x^\nu-q x^\nu  x^\mu=0   ~~~~~~~~&{\mbox{\sl quantum (hyper)plane}}
\end{eqnarray}
where $\theta^{\mu\nu}$ (a real antisymmetric matrix) ,
$f^{\mu\nu}_{~~\sigma}$ (real structure constants) , $q$ (a complex
number, e.g. a phase) are the respective
  noncommutativity parameters.
Quantum groups and quantum spaces \cite{FRT, Manin} are usually
described in this way.
In this case we do not have a space (i.e. a set of points), rather we have
a noncommutative algebra generated by the coordinates $\{x^\mu\}$ and
their relations; when the noncommutativity parameters ($\theta^{\mu\nu}, f^{\mu\nu}_{~~\sigma}, q$) are
turned off this algebra becomes commutative and is the algebra of
functions on a usual space. 
Of course we can also impose further constraints, for example
periodicity of the coordinates describing the canonical noncommutative
spacetime  (\ref{uuno}) (that typical of phase-space quantum mechanics) leads to a
noncommutative torus rather than to a noncommutative
(hyper)plane. Similarly, constraining the coordinates of the  quantum (hyper)plane relations 
(\ref{ttre}) we obtain a quantum (hyper)sphere. 
\sk
This algebraic description should then be complemented by a topological approach.
One that for example leads to the notions of continuous
functions. This is achieved completing the algebra
generated by the noncommutative coordinates to a
$C^\star$-algebra. Typically $C^\st$-algebras arise as algebras of
operators on Hilbert space. Connes noncommutative geometry \cite{connes} starts from
these notions and enriches the $C^\st$-algebra structure and its representation
on Hilbert space so to generalize to the noncommutative case also the
notions of smooth functions and metric structure. 
\sk
Another approach is the  $\star$-product one. Here we retain the usual
space of functions from  commutative space to complex numbers, but we deform the pointwise product operation in a
$\star$-product one. A $\star$-product sends two functions $(f,g)$ in
a third one $(f\star g)$. 
It is a differential operator on both its arguments (hence it is
frequently called a  bi-differential operator). It has the
associative property $f\st (g\st h)= (f\st
g)\st h$. 
The  most known example is the Gronewold-Moyal-Weyl star
product on ${\mathbb{R}}^{2n}$, 
\eq\label{GMW}
( f\st h)(x)=
{\rm e}^{\frac{i}{2}\theta^{\mu\nu}{\partial\over\partial x^\mu}\otimes
{\partial\over\partial y^\nu}}f(x)h(y)\big|_{x=y}~.
\en

Notice that  if we set 
\[\FF^{-1}={\rm e}^{\frac{i}{2}\theta^{\mu\nu}{\partial\over\partial x^\mu}\otimes
{\partial\over\partial y^\nu}}
\]
then
\[
( f\st h)(x)=
\mu\circ\FF^{-1}( f\otimes h)(x)
\]
where $\mu$ is the usual product of functions $\mu(f\otimes g)=fh$.
The  element 
$\FF={\rm e}^{-\frac{i}{2}\theta^{\mu\nu}{\partial\over\partial x^\mu}\otimes
{\partial\over\partial y^\nu}}
$
 is an example of a Drinfeld twist. It is defined by the  exponential series in powers
 of the noncommutativity parameters $\theta^{\mu\nu}$,
\eq
\FF={\rm e}^{-\frac{i}{2}\theta^{\mu\nu}{\partial\over\partial x^\mu}\otimes
{\partial\over\partial y^\nu}}
=1\otimes 1 -{i\over 2}\theta^{\mu\nu}\partial_\mu\otimes \partial_\nu
-{1\over 8}  \theta^{\mu_1\nu_1}\theta^{\mu_2\nu_2}
\partial_{\mu_1}\partial_{\mu_2}\otimes \partial_{\nu_1} \partial_{\nu_2} + \ldots\nonumber
\en
It is easy to see that $x^\mu\star x^\nu-x^\nu\star
x^\mu=i\theta^{mu\nu}$ thus also in this approach we recover the
noncommutative algebra (\ref{uuno}).

\sk
In this paper noncommutative spacetime will be spacetime equipped with
a $\star $-product. We will not discuss when the exponential series
$f\star g=fg-{i\over  2}\theta^{\mu\nu}\partial_\mu(f)\partial_\nu(g)+\ldots$ defining the
function $f\star g$ is actually convergent. We will therefore work in
the well established context of formal deformation quantization
\cite{Kontsevich}.  
In the latter part of the paper we will consider a series expansion of the
noncommutative gravity action in powers of the
noncommutativity parameters $\theta$, we will present the first
order in $\theta$ (a second order study appears in \cite{ACpreparation}), therefore the convergence aspect
won't be  relevant.


The method of constructing $\star$-products using twists is not the
most general method, however it is quite powerful,  and the class of
$\star$-products obtained is quite wide. For example choosing the
appropriate twist we can obtain 
the noncommutative relations (\ref{uuno}), (\ref{ddue}) and also (depending on the
structure constant explicit expression)  some of the Lie
algebra type (\ref{ttre}).

\section{Twists and $\st$-Noncommutative Manifolds
}
Let $M$ be a smooth manifold, a twist is an invertible element $\FF\in U\Xi\otimes U\Xi$
where $U\Xi$ is the universal enveloping algebra of vector fields, (i.e. it
is the algebra generated by vector fields on $M$ and where the element
$XY-YX$ is identified with the vector field $[X,Y]$). The element $\FF$
must satisfy some further conditions that we do not write here, but
that are satisfied if we consider abelian twists, i.e., twists of the form
\eqa
{\mathcal F^{}}&=&{\rm e}^{-\frac{i}{2}\theta^{IJ}
{X_I}\otimes
{X_J}}\nonumber\\
&=&1\otimes 1 -{i\over 2}\theta^{IJ}X_I\otimes X_J
-{1\over 8}  \theta^{I_1J_1}\theta^{I_2J_2}
X_{I_1}X_{I_2}\otimes X_{J_1} X_{J_2} + \ldots\nonumber
\ena
were the vector fields $X_I$ ($I=1,...s$ with $s$ not necessarily equal
to $m=\rm{dim} \,M$) are mutually commuting $[X_I,X_J]=0$ (hence
the name abelian twist).

It is convenient to introduce the following notation
\eqa
\mathcal{F}^{-1}
&=&1\otimes 1 +{i\over 2}\theta^{IJ}X_I\otimes X_J
-{1\over 8}  \theta^{I_1J_1}\theta^{I_2J_2}
X_{I_1}X_{I_2}\otimes X_{J_1} X_{J_2} +\ldots\nonumber\\
&=&\of^\alpha\otimes\of_\alpha\label{2.3}
\ena
where a sum over the multi-index $\alpha$ is understood.

\sk
Let $\A$ be the algebra of smooth functions on the manifold $M$. Then,
given a twist $\FF$, we deform $\A$ in a noncommutative algebra
$\A_\st$ by defining the new product of functions
$$f\st h=\of^\al(f)\,\of_\al(h)\,,$$
we see that this formula is a generalization of the Gronewold-Moyal-Weyl star
product on ${\mathbb{R}}^{2n}$ defined in (\ref{GMW}).
Since the vector fields $X_I$ are mutually commuting then this
$\star$-product is associative.
Note that only the algebra structure of $\A$ is changed to $\A_\st$
while, as vector spaces, $\A$ and $\A_\st$ are the same.
We similarly consider the algebra of exterior forms $\Omega^\bullet$ with the wedge
product $\wedge$, and deform it in the noncommutative
exterior algebra $\Omega_\st^\bullet$ that is characterized by the graded
noncommutative exterior product $\wedge_\st$  given by
$$\tau\wedge_\st \tau'=\of^\al(\tau)\wedge \of_\al(\tau')~,$$
where $\tau$ and $ \tau'$ are arbitrary exterior forms. Notice that
the action of the twist on $\tau$ and $\tau'$ is via the Lie
derivative: each vector field  $X_{I_1}, X_{I_2}, X_{J_1},
X_{J_2}\ldots$ in (\ref{2.3}) acts on forms via the Lie derivative.

It is not difficult to show that the usual exterior derivative is compatible with
the new $\wedge_\st$-product,
\eq
d(\tau\wedge_\star \tau')=d(\tau)\wedge_\star \tau'+(-1)^{deg(\tau) } \tau\wedge_\star
d\tau'
\en
this is so because the Lie derivative commutes
with the exterior derivative.

We also have compatibility with the usual undeformed integral (graded cyclicity property):
        \eq
       \int \tau \wedge_\star \tau' =  (-1)^{deg(\tau) deg(\tau')}\int \tau' \wedge_\star \tau\label{cycltt'}
       \en
(the equality holds up to boundary terms).
Finally we also have compatibility with the undeformed complex conjugation:
\eq
       (\tau \wedge_\star \tau')^* =   (-1)^{deg(\tau) deg(\tau')} \tau'^* \wedge_\star \tau^*~.
\en
{\bf Note.} We remark that all these properties are due to the special nature of
the $\star$-product we consider. As shown in \cite{Kontsevich}
$\star$-products are in 1-1 correspondence with Poisson structures 
$\{~,~\}$ on the manifold $M$. The Poisson structure the twist $\FF$ induces
is $\{f,g\}=i\theta^{IJ}X_I(f)X_J(g)$. However the twist $\FF$ encodes
more information than the Poisson bracket $\{~,~\}$. The key point 
is that the twist is associated with the Lie algebra $\Xi$ (and morally
with the diffeomorphisms group of $M$). Given a twist we can deform
the Lie algebra $\Xi$ (the diffeomorphisms group of $M$) and then,
using the Lie derivative action  (the action of the diffeomorphisms
group) we induce noncommutative deformations of the algebra of
functions on $M$, of the exterior algebra and more generally of the differential
and Riemannian geometry structures on $M$ \cite{wessgroup}, leading to
noncommutative Einstein equations for the metric tensor \cite{wessgroup}.

\sect{Noncommutative vierbein gravity coupled to fermions}
\subsection{Classical action}
The usual action of first-order gravity coupled to spin ${1 \over 2}$
fields reads:
 \eq
   S =\epsi_{abcd} \int  R^{ab} \we V^c \we V^d  -
i \bar\psi  \ga^a V^b \we V^c \we V^d \we D\psi 
-i   (D \bar\psi)   \ga^a \we V^b \we V^c \we V^d \psi
  \label{action1comp}
\en
\noi with
$
 R^{ab} = d\om^{ab}  - \om^a_{~c} \we \om^{cb}\,,
$ and  the Dirac conjugate 
defined as usual: $\psibar = \psi^\dagger \ga_0$. 
This action can be recast
in an index-free form \cite{Chamseddine}, \cite{AC1}, convenient for generalization to the
noncommutative case:
\eq
 S =  \int Tr \left(i R \we V \we V \ga_5\right)+\psibar   V \we V \we V \ga_5  D\psi +
 D\psibar \we V \we V \we V \ga_5 \psi \label{action1}
\en
\noi where 
 \eq
  R= d\Om - \Om \we
\Om, ~~~~~ D\psi = d\psi - \Om \psi,~~~~~D \psibar =\overline{D\psi}= d \psibar + \psibar  \Om    \label{psipsi}
\en
\noi with
$$ \Om  \equiv {1 \over 4} \om^{ab} \ga_{ab}, ~~~~~V  \equiv V^a \ga_a ,~~~~~R \equiv {1\over4} R^{ab} \ga_{ab}
$$ taking value in Dirac gamma matrices.
Use of the gamma matrix  identities
$
\ga_{abc} = i \epsi_{abcd} \ga^d \ga_5,$ $
 Tr (\ga_{ab} \ga_c \ga_d \ga_5) = -4 i \epsi_{abcd}
$
 in computing the trace leads back to the usual action (\ref{action1comp}).

\sk
The action (\ref{action1}) is invariant under local diffeomorphisms
 (because it is the integral of a 4-form on a 4-manifold)
  and under  local Lorentz rotations. 
In the index-free form they read
\eq
\de_\epsi V = -[V,\epsi ] , ~~~\de_\epsi \Om = d\epsi - [\Om,\epsi],~~~~
\de_\epsi \psi = \epsi \psi, ~~~\de_\epsi \psibar = -\psibar \epsi
\en
\noi with
$
 \epsi = {1\over 4} \epsi^{ab} \ga_{ab}
$
The local Lorentz invariance of the index free action follows from
$ \de_\epsi  R = - [{ R},\epsi ]$ and $\de_\epsi D\psi =
\epsi D\psi$, the cyclicity of the trace $Tr$ and the fact that the
gauge parameter $\epsi$ commutes with
   $\ga_5$.

\subsection{Noncommutative gauge theory and Lorentz group}

Consider an infinitesimal gauge transformation $\la=\la^AT^A$, where the
generators $T^A$ belong to some representation (the fundamental, the
adjoint, etc.) of a Lie group $G$. Since two consecutive gauge
transformations are a gauge transformation, the commutator of two
infinitesimal ones $[\la,\la']$, closes in the Lie algebra of $G$.
This in general is no more the case in noncommutative gauge theories.
In the noncommutative case the commutator of two infinitesimal gauge
transformations is
$$[\la\,\star\!\!\!_,~\la']\equiv\la\star \la'-\la'\star\la=
{1\over 2}\{\la^A\,\star\!\!\!_,~\la'^B\}\,[T^A,T^B]+{1\over
2}[\la^A\,\star\!\!\!_,~\la'^B]\,\{T^A,T^B\}~.$$
We see that also the anticommutator $\{T^A,T^B\}$ appears. This is
fine if our gauge group is for example $U(N)$ or $GL(N)$ in the fundamental or in the
adjoint, since in this case $\{T^A,T^B\}$ is again in the Lie
algebra, however for more general Lie algebras (including all simple
Lie algebras)  we have to enlarge the Lie
algebra to include also anticommutators besides commutators, i.e. we
have to consider all possible products $T^AT^B\ldots T^C$ of generators. 


\sk
Our specific case is the  Lorentz group in the spinor
representation given by the Dirac gamma matrices $\gamma_{ab}$. 
The algebra generated by these gamma matrices is that of all even
$4\times 4$ gamma matrices. 
The noncommutative gauge parameter will therefore have components
$$\epsi={1\over
  4}\epsi^{ab}\gamma_{ab}+i\epsi1\!\!1+\tilde\epsi\gamma_5~.$$ 
The extra gauge parameters $\epsi,\tilde\epsi$ can be chosen to
be real (like
$\epsi_{ab}$). Indeed the reality of 
$\epsi_{ab}$, $\epsi,\tilde\epsi$ is equivalent to the
hermiticity condition
\eq\label{hermconde}
 -\ga_0 \epsi \ga_0 =
 \epsi^\dagger
\en
and if the gauge parameters $\epsi$, $\epsi'$ satisfy this condition then also $ [\epsi\,\star\!\!\!_,~\epsi']$
is easily seen to satisfy this
hermiticity condition.

We have centrally extended the Lorentz group to
$$
SO(3,1)\rightarrow SO(3,1)\times U(1)\times R^{+}~,$$ or more precisely,
(since our manifold $M$ has a spin structure and we have a gauge
theory of the spin group $SL(2,C)$)
\[
SL(2,C)\rightarrow GL(2,C)\,.
\]
The Lie algebra generator $i1\!\!1$ is the anti-hermitian generator corresponding to
the $U(1)$ extension, while 
$\gamma_5$ is the hermitian generator  corresponding to the noncompact
$R^+$ extension. 
\sk

Since under noncommutative gauge transformations we have
\eq
\de_\epsi \Om = d\epsi - \Om \star \epsi+ \epsi
\star \Om
\label{stargauge}
 \en
also the spin connection and the curvature
will  be valued in the
$GL(2,C)$ Lie algebra representation given by all the even gamma matrices,
 \eq
  \Om = {1 \over 4} \om^{ab} \ga_{ab} + i \om 1\!\!1 + \omtilde \ga_5,
  ~~~~~
 R= {1\over 4} R^{ab} \ga_{ab} + i r 1\!\!1 + \rtilde \ga_5~.
\en
Similarly the gauge transformation of the vierbein,
\eq\label{stargauge2}
\de_\epsi V = -V \star \epsi + \epsi \star V,
 \en
closes in the vector space of odd gamma matrices (i.e. the vector
space linearly generated by $\gamma^a,\gamma^a\gamma_5$)
and not in the subspace of just the $\gamma^a$ matrices.
Hence the noncommutative vierbein are  valued in the odd gamma matrices
\eq
V = V^a
\ga_a + \Vtilde^a \ga_a \ga_5  ~.
  \en

\subsection{Noncommutative Gravity action and its symmetries}\label{4.3}
We have all the ingredients in order to generalize to the noncommutative case
the gravity action coupled to spinors of Section 3: 
an abelian twist giving the star products of functions and forms on
the spacetime manifold
$M$ (and compatible with usual integration on $M$);
an extension to $GL(2,C)$ of the Lorentz gauge group, so that
infinitesimal noncommutative gauge transformations close in this
extended Lie algebra.
The action reads
\eq
 S =  \int Tr \left(i {R}\westar \! V \!\westar\! V \ga_5\right) \,+\,
\bar\psi  \star V \!\westar \! V\! \westar \!V\!\westar \!\gamma_5D\psi 
\,+\,  D \bar\psi \westar V \!\westar \! V\! \westar \!V\!\star
\gamma_5\psi
\label{action1NC}
\en
\noi with
 \eq R= d\Om - \Om \westar \Om~,~~~~ D\psi = d\psi -
\Om \star \psi ~,~~~~D \psibar = d \psibar + \psibar  \star \Om    \label{2.20}~.
\en

\noi{\bf Gauge invariance}\\
The invariance of the noncommutative action (\ref{action1NC}) under
the $\star$-variations is demonstrated  in exactly the same way as 
for the commutative case: noting that besides (\ref{stargauge}) and
(\ref{stargauge2}) we have
   \eq
\de_\epsi \psi = \epsi \star \psi,
~~~\de_\epsi \psibar = -\psibar \star \epsi
\label{stargauge3}~,~~
  \de_\epsi D\psi = \epsi \star D\psi~,~~
\de_\epsi D\psibar = -D\psibar \star \epsi~,~~
  \de_\epsi R = - R \star \epsi+ \epsi \star R~,
   \en
 and using that
   $\epsi$  commutes with $\ga_5$, and  the cyclicity of the
   trace together with the graded cyclicity of the integral with respect to the $\star$-product.  
\sk
\noi{\bf Diffeomorphisms invariance} \\
The $\star$-action (\ref{action1NC}) is invariant under usual diffeomorphisms,
being the integral of a $4$-form. Under these diffeomorphisms the
vector fields $X_I$ transform covariantly.  We also mention that since
the vector fields $X_I$ appear only in the $\star$-product,  the action
is furthermore invariant under $\star$-diffeomorphisms as defined in \cite{wessgroup}, see
discussion in \cite{book}, Section 8.2.4. Under these deformed diffeomorphisms the vector fields $X_I$ do not
transform.
\sk
\noi
{\bf{Reality of the action~}}\\
 Hermiticity conditions can be imposed on the fields $V$ and $\Om$ as
 done with the gauge parameter $\epsi$ in (\ref{hermconde}) :
\eq
 \ga_0 V \ga_0 = V^\dagger,~~~ -\ga_0 \Omega \ga_0 =
 \Omega^\dagger.~
\label{hermconjNC}
 \en
\noi These hermiticity conditions are consistent with the gauge
variations and can be used to check that the action (\ref{action1NC}) is
real by comparing it to its complex conjugate (obtained by taking the Hermitian conjugate of the
4-form inside the trace in the integral).
As previously observed for
the component gauge parameters $\epsi^{ab}$, $\epsi$,  $\epsitilde$, the hermiticity conditions (\ref{hermconjNC}) imply that the component fields
$V^a$, $\Vtilde^a$, $\om^{ab}$, $\om$, and $\omtilde$ are real fields.
\sk
\noi {\bf Charge conjugation invariance}\\
Noncommutative charge conjugation is the following
transformation (extended linearly and multiplicatively\footnote{Multiplicativity and the transformation $\star\to \star^C_\theta=\star_{-\theta}$ is equivalent to antimultiplicativity with respect to the $\star$-product: $(f\star g)^C=g^C\star f^C$ (for $f$ and $g$ scalar fields that are not matrix valued).}):
\eq\label{defCconj}
\psi\to\psi^{\;C}\equiv C(\bar\psi)^T=-\gamma_0 C \psi^\ast\!~,~~
V\to V^{\,C}\equiv
C{\,V}^{\;T}C\!~,~~\Omega\to \Omega^{\,C}\equiv 
C{\:\!
\Omega_{}}^{\;T}C\!~,~~\star_\theta\to\star_\theta^C=\star_{-\theta}~
\!~,
\en
and consequently $\wedge_{\star_\theta}\to\wedge_{\star_\theta}^{\,C}=\wedge_{\star_{-\theta}}~.$ 
Then the action (\ref{action1NC}) is invariant under charge
conjugation. For example
\eqa S^C_{bosonic}\!\!&=& i  \int Tr ( {R^C}\wedge_{-\theta} V^C
\wedge_{-\theta} V^C \ga_5 )^T =
- i  \int Tr ( {R^T}\wedge_{-\theta} V^T \wedge_{-\theta} V^T
C\ga_5C^{-1} )^T \nonumber\\
&=& -i \int Tr 
 \left( (V^T \we_{-\theta} V^T \ga_5^T)^T \we_\star R \right)  
= -i \int Tr \left( -(V^T \ga^T_5)^T \we_\star V \we_\star R \right) \nonumber \\ 
& =& i \int Tr ( \ga_5
 V \we_\star  V  \we_\star R) = i \int Tr (R\we_\star  \ga_5
 V \we_\star  V) 
 =   i \int Tr (R\we_\star  
 V \we_\star  V \ga_5) \nonumber \\ 
&=&S_{bosonic}
\ena
Similarly the fermionic part of the action satisfies
$S_{fermionic}^C=S_{fermionic}$. 

\sk
\noi
{\bf Charge conjugation conditions }\\
In the classical limit $\theta\rightarrow 0$ the $\star$-product
becomes the usual pointwise product. The noncommutative gauge symmetry
becomes a usual gauge symmetry with gauge group $GL(2,C)$ and the
noncommutative vierbein in the classical limit leads to two
independent vierbeins: $V^a$ and
$\tilde V^a$ transforming both only under the $SL(2,C)$ subgroup of
$GL(2,C)$. As observed in \cite{Chamseddine} this is problematic because we obtain two massless
gravitons, and only one local Lorentz symmetry, that is  not enough in
order to kill the unphysical degrees of freedom. Either we concoct a
mechanism such that the second
graviton becomes massive or we further constraint the noncommutative
theory so that in the classical limit the extra vierbein vanishes.

The vanishing of the $\tilde V^a$ components in the classical limit is
achieved by imposing charge conjugation conditions on the fields \cite{AC1}:
 \eq
 C V_\theta (x) C = V_{-\theta} (x)^T,~~~C \Omega_\theta (x) C = \Omega_{-\theta} (x)^T,~~~
 C \epsi_\theta (x) C = \epsi_{-\theta} (x)^T \label{ccc}
 \en
These conditions  involve the $\theta$-dependence of the fields. This latter
is due to  the $\star$-product $\theta$-dependence (recall that the $\star$-product is defined as an
expansion in power series of the noncommutativity parameter
$\theta$). 
Since  noncommutative gauge transformations involve the
$\star$-product, the gauge transformed fields will be
$\theta$-dependent and hence field configurations  are in general
$\theta$-dependent.

Conditions (\ref{ccc}) are consistent with the
$\star$-gauge transformations. For example the field $ C V_\theta (x)^T C$ can
be shown to transform in the same way as $V_{-\theta} (x)$ \cite{AC1}.
\sk
\noi For the component fields and  gauge parameters the charge
conjugation conditions imply that the components $V^a,  \om^{ab}$ are
even in $\theta$, while the components  $\Vtilde^a, \om^{},\omtilde$
are odd:
 \eqa & & V^a_\theta=V^a_{-\theta}, ~~~
\om^{ab}_\theta = \om^{ab}_{-\theta} \\ & & \Vtilde^a_\theta
=-\Vtilde^a_{-\theta}, ~~~ \om^{}_\theta=- \om^{}_{-\theta} ,~~~\omtilde^{}_\theta=
- \omtilde^{}_{-\theta}. \label{cconjonfields}
 \ena
 \noi Similarly for the gauge parameters:
$\epsi^{ab}_\theta= \epsi^{ab}_{-\theta}\,,~  \epsi^{}_\theta =- \epsi^{}_{-\theta} ,~~~\epsitilde^{}_\theta=-
\epsitilde^{}_{-\theta} $. In particular, since the components $\tilde V^a$ are odd in $\theta$ we
achieve their vanishing in the classical limit.
\sk
We can also conclude that the bosonic action is even in $\theta$.
Indeed  (\ref{ccc}) implies
\eq
V^{\,C}= 
V_{-\theta}~,~~
\Omega^{\,C}=\Omega_{-\theta}~,~~
R^{\,C}=R_{-\theta}~.~~
\en
Hence the bosonic action $S_{bosonic} (\theta) $ is mapped into
$S_{bosonic} (-\theta)$ under charge conjugation. Also for the fermionic
action, $S_{fermionic} (\theta)$, we have $S_{fermionic}
(\theta)^C=S_{fermionic} (-\theta)$ 
if the fermions are Majorana, i.e. if they satisfy 
$\psi^C=\psi_{-\theta}$.
{}From $S_{bosonic}(\theta)=S_{bosonic}(-\theta)$ we conclude that all
noncommutative corrections to the
classical action of pure gravity are even in $\theta$; this is also
the case if we couple noncommutative gravity  to Majorana fermions.

\section{Seiberg-Witten map (SW map)}
In the previous section we have formulated a noncommutative gravity
theory that in the classical limit $\theta\to 0$  reduces to usual
vierbein gravity. In the full noncommutative regime it has however a
doubling of the vierbein fields. We can insist on a noncommutative
gravity 
theory that has the same degrees of freedom of the classical one.
This is doable if we use the Seiberg-Witten map to express the
noncommutative fields in terms of the commutative ones. In this way
the gauge group is the Lorentz group $SL(2,C)$ and not the centrally
extended one $GL(2,C)$, indeed  also the  noncommutative gauge
parameters $\epsi^{ab},\epsi, \tilde\epsi$
are expressed in term of the commutative ones $\epsi^{ab}$.

Because of  the SW map the noncommutative fields can therefore be
expanded in terms of the commutative ones, and hence the noncommutative
gravity action can be  expanded, order by order in powers of $\theta$,
in terms of the usual commutative gravity and spinor field as well as
of the noncommutativity vector fields $X_I$. 
As we will see, we thus obtain a commutative gravity
action that at zeroth order in $\theta$ is usual gravity (coupled to
spinors) and at  higher orders in $\theta$ contains higher
derivative terms describing gravity and spinor fields coupled to the noncommutativity vector
fields $X_I$.
\sk
The Seiberg-Witten map (SW map) relates the noncommutative gauge
fields $\Omega$ to the ordinary $\Omega^0$, and the noncommutative gauge
parameters $\epsi$ to the ordinary $\epsi^0$ and $\Omega^0$ so as to satisfy:
 \eq
 \Omega(\Omega^0) + \de_\epsi \Omega (\Omega^0) = \Omega (\Omega^0 + 
\de_{\epsi^0} \Omega^0) \label{SWcondition}
 \en
 with the noncommutative and ordinary gauge variations given by 
  \eq
  \de_\epsi \Omega= \dd\epsi -  \Omega
      \star\epsi +\epsi \star\Omega~~,~~~  \de_{\epsi^0} \Omega^0= \dd \epsi^0 
-  \Omega^0
      \epsi^0 +\epsi^0\Omega^0~.
\label{3.44h}
      \en
Equation  (\ref{SWcondition}) can be solved order by order in $\theta$
\cite{Ulker}, yielding
$\Omega$ and $\epsi$ as  power series in $\theta$:  
 \eqa 
    \Omega(\Omega^0, \theta) &=& \Omega^0 +\Omega^1 (\Omega^0)  + \Omega^2 (\Omega^0) + \cdots + \Omega^n (\Omega^0)+ \cdots  \\
     \epsi (\epsi^0, \Omega^0, \theta)  &=&  \epsi^0 + \epsi^1 (\epsi^0, A^0)+ \epsi^2 (\epsi^0, \Omega^0)+ \cdots + 
     \epsi^n (\epsi^0, \Omega^0)+ \cdots 
     \ena    
 \noi  where $\Omega^n (\Omega^0)$ and $\epsi^n (\epsi^0, \Omega^0)$  are of order $n$ in $\theta$. Note that  $\epsi$
depends on the ordinary $\epsi^0$ and also on $\Omega^0$.
\sk
The Seiberg-Witten condition (\ref{SWcondition}) states that the dependence of the noncommutative gauge field on the ordinary one is fixed
by requiring that ordinary gauge variations of $\Omega^0$ inside
$\Omega(\Omega^0)$ 
produce the noncommutative gauge variation of $\Omega$. 
This implies that once we expand, order by order in $\theta$, the
noncommutative action in terms of the commutative fields, the
resulting action will be gauge invariant under ordinary local Lorentz
transformations because the noncommutative action is invariant under
the local noncommutative Lorentz transformations of Section \ref{4.3}.

Similarly  for matter fields we have 
$\psi(\psi^0,\Omega^0)+\delta_\epsi \psi(\psi^0,\Omega^0)=
\psi(\psi^0+\delta_{\epsi^0}\psi^0,\Omega^0+\delta_{\epsi^0}\Omega^0)
$ so that again noncommutative gauge variations correspond to
commutative ones.

Following ref. \cite{AC2}, up to first order in $\theta$ the solution
to the Seiberg-Witten conditions reads:
\eqa
& & 
\Omega=\Omega^0+\frac{i}{4}\theta^{IJ}\{\Omega^0_I\label{omsol}
, \ell_J \Omega^0+  R^0_{J} \}\\
& & \epsi= \epsi^0+ {i \over 4} \theta^{IJ} \{\Om^0_I, \ell_J \epsi^0 \} \\
 & & R = R^0+{i \over 4} \theta^{IJ} \left( \{\Om^0_I, (\ell_J + L_J)  R^0  \} 
 - [R^0_{I},R^0_{J} ] \right) \label{Rsol}\\
 & & \psi= \psi^0+{i \over 4} \theta^{IJ} \Om^0_I  (\ell_J + L_J) \psi^0 \label{solSW}
\ena
where
$\Om^0_A$, $R^0_A$ are defined as the contraction along the tangent
vector $X_I$ of
the exterior forms $\Om^0$, $R^0$, i.e. $\Om^0_A\equiv i_I\Om^0$,
$R^0_I \equiv i_I R^0$, ($i_I$ being the contraction along $X_I$).
We have also introduced the Lie derivative $\ell_I$ along the vector
field $X_I$, and the covariant Lie derivative $L_I$ along the
vector field $X_I$. $L_I$ acts on $R^0$ and $\psi^0$ as 
$L_I R^0 =\ell_I R^0-\Om^0_B \star
R^0+ R^0 \star\Om^0_I$ and
$L_I  \psi^0 = \ell_I \psi^0 - \Om^0_I \psi^0 $.
In fact the covariant Lie derivative $L_I$ has the Cartan form:
 \eq
  L_I = i_I D + D i_I~\nonumber
    \en
where $D$ is the covariant derivative.
We refer to \cite{AC2} for higher order in $\theta$ expressions.
All these formulae are {\it not} $SO(1,3) $-gauge covariant, due to
the presence of the ``naked" connection
$\Om^0$ and the non-covariant Lie derivative $\ell_I=i_I{\dd}+{\dd}
i_I$. 
However, when inserted in the NC action the resulting action is gauge invariant
order by order in $\theta$. Indeed usual gauge variations induce the
$\star$-gauge variations under which the NC action is
invariant. Therefore the NC action, re-expressed in terms of ordinary
fields via the SW map, is invariant under usual gauge
transformations. Since these do not involve $\theta$, the expanded
action is invariant under ordinary gauge variations order by order in
$\theta$. Moreover the action, once re-expressed in terms of ordinary fields remains geometric,
and hence invariant under diffeomorphisms. This is the case because the noncommutative
action and the SW map are geometric, we indeed see that only
coordinate independent operations like the contraction $i_I$ and the
Lie derivatives $\ell_I$ and $L_I$ appear in the SW map. 

{}From (\ref{omsol}) and (\ref{solSW}) we also deduce
\eq
D\psi= D\psi^0+\frac{i}{4} \theta^{IJ} \Big(\Om^0_I (\ell_J + L_J) D\psi^0
-2R_I L_J\psi\Big)\label{solSWDpsi}~.
\en

\section{Action at first order in $\theta$}
The expression of the gravity action, up to second order in $\theta$,
in 
terms of the commutative fields and of the first order fields
 (\ref{omsol})-(\ref{solSWDpsi}) has been given in \cite{AC2}.
The action is gauge invariant even if the expression in \cite{AC2} is
not explicitly gauge invariant.
We here present the explicit  gauge invariant expression for the
action up to first order in $\theta$.
We replace the noncommutative fields appearing in the action
with their expansions (\ref{omsol})-(\ref{solSWDpsi}) in commutative fields 
and integrate by parts in order to obtain
an explicit $SO(3,1)$ gauge invariant action.  We thus obtain
the following gravity action coupled to spinors
\eqa\label{action11}
S &\!\!=\!\!&\!\!  \int Tr \left(i R V V \ga_5\right)+\psibar   V^3\ga_5  D\psi +
 D\psibar V^3 \ga_5 \psi \\ 
&
&+\frac{i}{4}\theta^{IJ}\Big(\bar\psi\{V^3,R_{IJ}\}\gamma_5D\psi+D\bar\psi\{V^3,R_{IJ}\}\gamma_5\psi\Big)\nonumber\\[.5em]
& &+\frac{i}{2}\theta^{IJ}\Big(2L_I\bar\psi
R_JV^3\gamma_5\psi-2\bar\psi V^3R_I\gamma_5L_J\psi - L_I\bar\psi V^3\gamma_5L_JD\psi-L_ID\bar\psi\,
V^3\gamma_5 L_J\psi \nonumber\\
&&~~+\bar\psi (\{L_I\!VL_J\!V,V\}+L_I\!V\, V L_J\!V)\gamma_5D\psi +D\bar\psi  (\{L_I\!VL_J\!V,V\}+L_I\!V\, V L_J\!V)\gamma_5\psi 
\Big)
+O(\theta^2)\nonumber
\ena
where with obvious abuse of notation we have omitted the apex ${}^0$ denoting
commutative fields, we also have omitted writing the wedge product, and
$V^3=V\wedge V\wedge V$.

\sk
\section{Conclusions}
We have constructed an extended Einstein gravity action that:
{\sl i)} is explicitly invariant under local Lorentz transformations because 
expressed solely in terms of the gauge covariant operators $L_I, i_I,
D$ and fields $R, V, \psi\:\!$; 
{\sl ii)} is diffeomorphic invariant;
{\sl iii)} has the same fields of classical gravity plus the noncommutative
structure. This latter is given by the vector fields $\{X_I\}$ that choosing
an appropriate kinetic term can become dynamical, the idea being that
both spacetime curvature and noncommutativity should depend on  matter
distribution.

This extended action has been obtained from considering gravity on
noncommutative spacetime, that as argued in the introduction is a very
natural assumption at high energies, like those close to the
inflationary epoch.
\sk\sk
\noi {\large \bf Acknowledgements}\\
We acknowledge the fruitful and pleasant atmosphere
and the perfect organization  enjoyed during the Udine Symposium.  
\appendix
\sect{Gamma matrices in $D=4$}

We summarize in this Appendix our gamma matrix conventions in $D=4$.

\eqa
& & \eta_{ab} =(1,-1,-1,-1),~~~\{\ga_a,\ga_b\}=2 \eta_{ab},~~~[\ga_a,\ga_b]=2 \ga_{ab}, \\
& & \ga_5 \equiv i \ga_0\ga_1\ga_2\ga_3,~~~\ga_5 \ga_5 = 1,~~~\epsi_{0123} = - \epsi^{0123}=1, \\
& & \ga_a^\dagger = \ga_0 \ga_a \ga_0, ~~~\ga_5^\dagger = \ga_5 \\
& & \ga_a^T = - C \ga_a C^{-1},~~~\ga_5^T = C \ga_5 C^{-1}, ~~~C^2 =-1,~~~C^\dagger=C^T =-C
\ena

\end{document}